\documentclass[aip,cha,reprint,floatfix]{revtex4-1}

\usepackage{amsmath}
\usepackage{amsfonts}
\usepackage{color}
\usepackage{graphicx}
\usepackage{times}
\usepackage{ifpdf}

\newcommand{\thi}{\ensuremath{\theta_i}}
\newcommand{\thr}{\ensuremath{\theta_r}}

\begin{document}

\title{Non-specular reflections in a macroscopic system with
  wave-particle duality: spiral waves in bounded media. }


\affiliation{Mathematics Institute, University of Warwick, Coventry CV4 7AL, UK}
\author{Jacob Langham}
\email[]{J.Langham@warwick.ac.uk}
\author{Dwight Barkley}
\email[]{D.Barkley@warwick.ac.uk}

\date{\today}

\begin{abstract}
Spiral waves in excitable media possess both wave-like and particle-like
properties. When resonantly forced (forced at the spiral rotation frequency)
spiral cores travel along straight trajectories, but may reflect from medium
boundaries.  
Here, numerical simulations are used to study reflections from two types
of boundaries.  The first is a no-flux boundary which waves cannot cross, while
the second is a step change in the medium excitability which waves do cross.
Both small-core and large-core spirals are investigated.  The predominant
feature in all cases is that the reflected angle varies very little with
incident angle for large ranges of incident angles. Comparisons are made to
the theory of Biktashev and Holden. Large-core spirals exhibit other phenomena
such as binding to boundaries.  The dynamics of multiple reflections is
briefly considered.
\end{abstract}


\maketitle



\begin{quotation}
Wave-particle duality is typically associated with quantum mechanical
systems. However, in recent years it has been observed that some macroscopic
systems commonly studied in the context of pattern formation also exhibit
wave-particle duality.  Two systems in particular have attracted considerable
attention in this regard: drops bouncing on the surface of a vibrated liquid
layer~\cite{coud05,coud052,prot06,coud06,eddi09,eddi092} and waves in chemical
media~\cite{bikta03,Biktashev:2007,Sakurai:2002,Steele:2008,Biktasheva:2010,
Biktashev:2010}.
The second case is the focus of this paper.
We explore non-specular reflections associated with spiral waves in excitable
media -- reflections not of the waves themselves, but of the particle-like
trajectories tied to these waves.
\end{quotation}
\section{Introduction}
\begin{figure}[ht]
\centering
\includegraphics[width=3.25in]{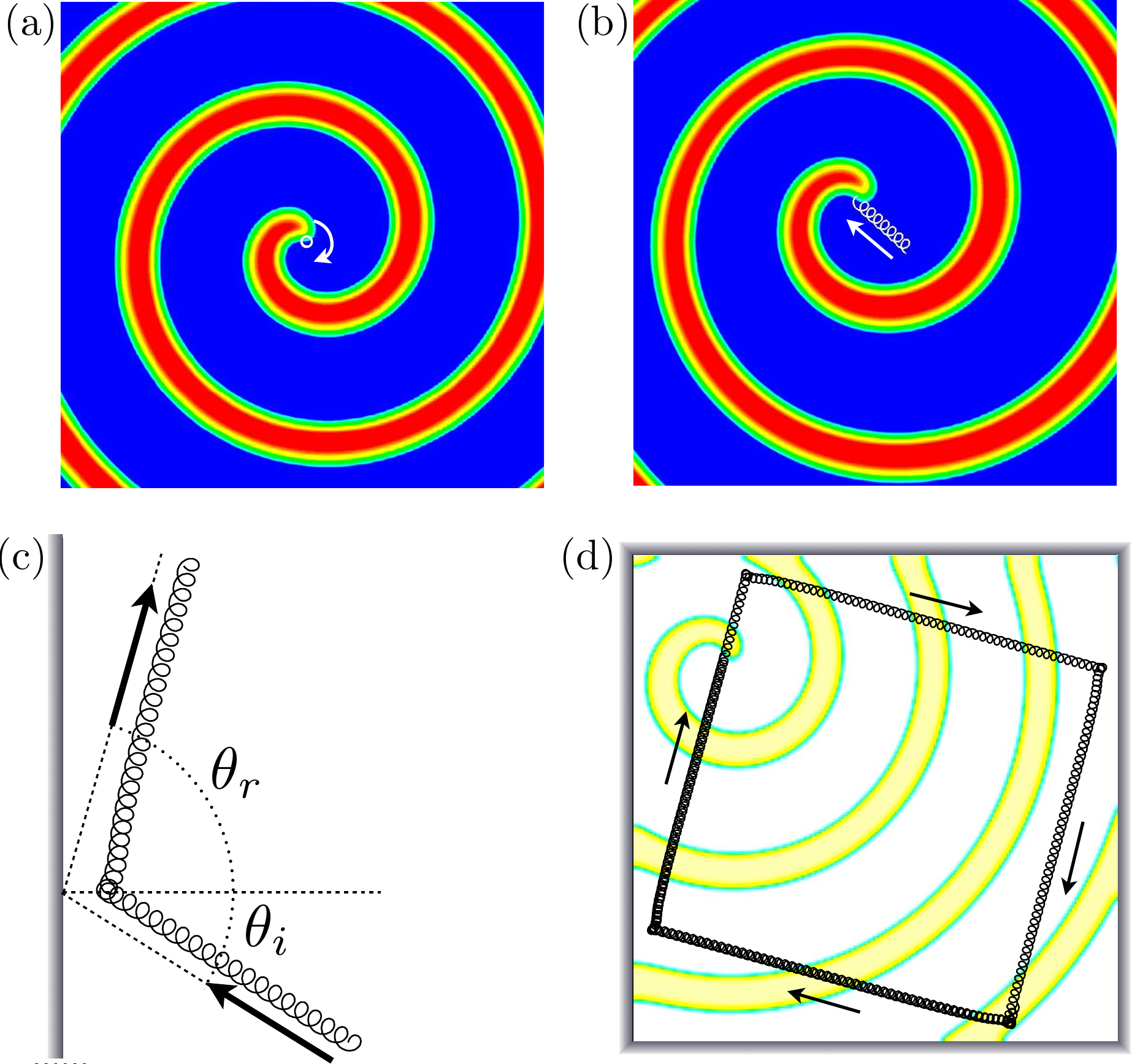} 
\caption{Illustration of resonant drift and reflection for spiral waves in
excitable media. (a) Periodically rotating spiral wave in the unforced
regime. The wave rotates around a fixed core and the path of the spiral tip
(white) is a circle. (b) Resonant drift. The medium is parametrically forced
at the spiral rotation frequency. The core moves along a straight path and the
spiral tip traces out a cycloid (white). (c) Reflection of drifting spiral
from a no-flux boundary. The incident and reflected angles, \thi\ and \thr,
are indicated. (d) Path of a drifting spiral in a square box. The underlying
spiral wave at one instant in time is shown faintly.
In all cases the plotted fields are the excitation variable, $u$, of the
reaction diffusion model. Details are given later in the text.}
\label{fig:intro} 
\end{figure}

Rotating spirals are a pervasive
feature of two-dimensional excitable media, such as the Belousov-Zhabotinsky
reaction~\cite{Belo59,Zhabotinsky:1964,Zhab70,Winf84}.
Figure~\ref{fig:intro}(a) illustrates a spiral wave from a standard model of
excitable media discussed below.  
The wave character of the system is evident. As the spiral rotates, a periodic
train of excitation is generated which propagates outward from the center, or
core, of the spiral.
Much of the historical study of excitable media has focused on the wave
character of the problem, as illustrated by efforts to determine the selection
of the spiral shape and rotation
frequency~\cite{Tyson:1988,Karma:1991,Bernoff:1991,Margerit:2001,Margerit:2002}.

However, it is now understood that these spiral waves also have particle-like
properties. This was first brought to the forefront by Biktasheva and
Biktashev~\cite{bikta03} and has been developed in more recent
years
\cite{Biktashev:2007,Biktasheva:2010,Biktashev:2010,Biktasheva:2009,
Biktashev:2011}.
One of the more striking illustrations of a particle-like property is {\em
resonant
drift}~\cite{Agladze:1987,Davydov:1988,bikt93,bikt95,stein93,schra95,mant96,zhan04,Biktashev:2007,Biktasheva:2010},
shown in Fig.~\ref{fig:intro}(b)-(d).
Resonant drift can occur spontaneously through instability, or due to spatial
inhomogeneity, or as here, by
means of resonant parametric forcing (periodically varying the medium
parameters in resonance with the spiral rotation frequency). 
As is seen, the core of the spiral drifts along a straight line. The speed is
dictated by the forcing amplitude while the direction is set by the phase of
the forcing, or equivalently the initial spiral orientation. 

The trajectories of drifting spirals are unaffected by the domain boundaries
(or other spirals should they be present) except on close approach, where
often the result is a reflection of the drifting core
\cite{bikt93,bikt95,Olmos:2008}, as illustrated in Fig.~\ref{fig:intro}(c).
Reflections are not specular -- the reflected angle \thr\ is not in general
equal to the incident angle \thi.
When placed in a square box, the drift trajectory typically will ricochet off
each boundary in such a way to eventually be attracted to a unique square path
where $\thi + \thr = 90^\circ$, as shown in figure~\ref{fig:intro}(d). (This
is the more common case, but others are considered herein.)

The primary goal of this paper is firstly to determine accurately, through
numerical simulations, the relationship between reflected and incident angles
for some representative cases of spiral waves in excitable media, and secondly
to explore the qualitative features of reflections in excitable media,
particularly multiple reflections in square domains.
While the numerical and theoretical study of reflecting trajectories was
undertaken by Biktashev and Holden many years ago~\cite{bikt93,bikt95}, 
much more extensive results are now possible and desirable, 
especially since phenomena strikingly similar to that seen in
Figs.~\ref{fig:intro}(c) and (d) have been observed in other macroscopic
systems with both wave-like and particle-like
properties~\cite{prot06,eddi092,Sakurai:2002,Steele:2008,
Prati:2011,ShowalterPC}.
%



\section{Model and Methods}

Our study is based on the standard Barkley model describing a generic
excitable medium~\cite{bark91}. In the simplest form the model is given by the
reaction-diffusion equations
\begin{align}
& \frac{\partial u}{\partial t} =  
\nabla^2u + \frac{1}{\epsilon}u(1-u)\left(u-\frac{v+b}{a}\right), \\
& \frac{\partial v}{\partial t} = u - v, 
\end{align}
where $u(x,y)$ is the excitation field (plotted in Fig.~\ref{fig:intro}) and
$v(x,y)$ is the recovery field; $a$, $b$, and $\epsilon$ are parameters.  
The parameters $a$ and $b$ collectively control the threshold for and duration
of excitation while the parameter $\epsilon$ controls the excitability of the
medium by setting the fast timescale of excitation relative to the timescale
of recovery.

We consider two parameter regimes -- known commonly as the small-core and
large-core regimes. The small-core case is shown in Fig.~\ref{fig:intro}. As
the name implies, the core region of the spiral, where the medium 
remains unexcited over one rotation period, 
is small. This is the more generic case for the Barkley model
and similar models and occupies a relatively large region of parameter space
in which waves rotate periodically. Small-core spirals are found in the lower
right part of the standard two-parameter phase diagram for the Barkley
model (see figure 4 of Ref.~\onlinecite{bark08}).
Large-core spirals rotate around relatively large regions 
[see Fig.~\ref{fig:large_core_intro}(a) discussed below].  Such
spirals occur in a narrow region of parameter space \cite{bark08,bark94} near
the boundary for propagation failure.  The core size diverges to infinity near
propagation failure.

Parametric forcing is introduced through periodic variation in the
excitability. Specifically we vary $\epsilon$ according to
\begin{equation}
\epsilon(t) = \epsilon_0 \left( 1 + A \sin(\omega_f t + \phi) \right),
\label{eq:forcing}
\end{equation}
where $A$ and $\omega_f$ are the forcing amplitude and frequency. The phase
$\phi$ is used to control the direction of resonant drift. The forcing
frequency producing resonant drift will be close to the natural, unforced,
spiral frequency. However, due to nonlinearity, there is a change in the
spiral rotation frequency under forcing and so $\omega_f$ must be adjusted
with $A$ to produce resonant drift along a straight line.

We have studied reflections in two situations. The first is reflection from a
no-flux boundary. This type of boundary condition corresponds to the wall of a
container containing the medium.  
We set the reflection boundary to be at
$x=0$ and impose a homogeneous Neumann boundary condition there:
\begin{equation}
\frac{\partial u}{\partial x}(0,y) = 0.
\label{eq:Neumann}
\end{equation}
Since there is no diffusion of the slow variable, no boundary condition is
required on $v$.  The medium does not exist for $x<0$.

The second situation we have studied is reflections from a step change
in excitability across a line {\em within} the medium.  We locate step change
on the line $x=0$.  We vary the threshold for excitation across this line by
having the parameter $b$ vary according to
\begin{equation}
b(x,y) = 
 \begin{cases} 
   b_0 & \text{if $x \ge 0$} \\
   b_0 - \triangle b & \text{if $x < 0$}
 \end{cases}
\label{eq:medium}
\end{equation}
Unlike for the no-flux boundary, in this case waves may cross the line $x=0$ and
so there is no boundary to wave propagation. Nevertheless, drifting spirals may
reflect from this step change in the medium and we refer to this a {\em
step boundary}.

The numerical methods for solving the reaction-diffusion equations are
standard and are covered elsewhere~\cite{bark91,Dowle:1997}. 
Some relevant computational details
particular to this study of spiral reflections are as follows. A converged
spiral for the unforced system is used as the initial condition. Simulations
are started with parametric forcing and the spiral drifts in a particular
direction dictated by the phase $\phi$ in Eq.~\eqref{eq:forcing}.  The
position of the spiral tip is sampled once per forcing period and from this
the direction of drift, i.e.\ the incident angle \thi, is determined by a
least-squares fit over an appropriate range of drift (after the initial spiral
has equilibrated to a state of constant drift, both 
in speed and direction, but
before the spiral core encounters a boundary). Likewise, from a fit to the
sampled tip path after the interaction with the boundary, we determine the
reflected angle \thr.
By varying $\phi$ we are able to scan over incident angles. 

The simulations are carried out in a large rectangular domain with no-flux
boundary conditions on all sides. For reflections from a Neumann
boundary~\eqref{eq:Neumann}, we simply direct waves to the computational
domain boundary corresponding to $x=0$. 
We also study reflections more globally from all
sides of a square domain with Neumann boundary conditions, 
such as in Fig.~\ref{fig:intro}(d). 
In the study of reflections from the step boundary~\eqref{eq:medium}, the
computational domain extends past the step change in parameter.  We have run
cases with the left computational boundary both at $x=-7.5$ and $x=-15$ and 
these are sufficiently far from $x=0$ that trajectory reflection is not
affected by the computational domain boundary. 
The dimensions of the rectangular computational domain are varied depending on
the angle of incidence. For $\thi \simeq \pm 90^\circ$ we require a long
domain in the $y$-direction, whereas for $\thi \simeq 0^\circ$ a much smaller
domain may be used.
In all cases we use a grid spacing of $h=1/4$. The time step is varied to
evenly divide the forcing period, but $\triangle t \simeq 0.019$ is typical.
Except where stated otherwise, the model parameters for the small-core case
are: $a = 0.8, b = 0.05$, and $\epsilon_0 = 0.02$. For the large-core case
they are $a = 0.6, b = 0.07$, and $\epsilon_0 = 0.02$. For the step boundary
$b_0 = 0.05$ and $\triangle b = 0.025$. Different values of the forcing
amplitude and period, $A$ and $\omega_f$, are considered.
Given the desire to measure incident and reflected angles precisely, we have
required drift be along straight lines to high precision and in
turn this has required high accuracy in the imposed forcing amplitude and
period. 
In the appendix we report the exact values for the forcing parameters used in
the quantitative incidence-reflection studies.


\section{Results}

Before presenting results from our study of reflections, it is important to be
precise about the meaning of incident and reflected angles. As is standard,
angles are measured with respect to the boundary normal. This is illustrated
in Fig.~\ref{fig:intro}(c). What needs to be stressed here is that spirals
have a chirality -- right or left handedness -- and this implies that we need
to work with angles potentially in the range $[-90^\circ, 90^\circ]$, rather
than simply $[0^\circ, 90^\circ]$.

Specifically, we consider clockwise rotating spirals and define \thi\ to be
positive in the clockwise direction from the normal. We define \thr\ to be
positive in the counterclockwise direction from the normal. Both
\thi\ and \thr\ are positive in Fig.~\ref{fig:intro}(c)
and for specular reflections $\thr=\thi$.  

\subsection{Small-core case}

\begin{figure}
\centering
\includegraphics[width=3.00in]{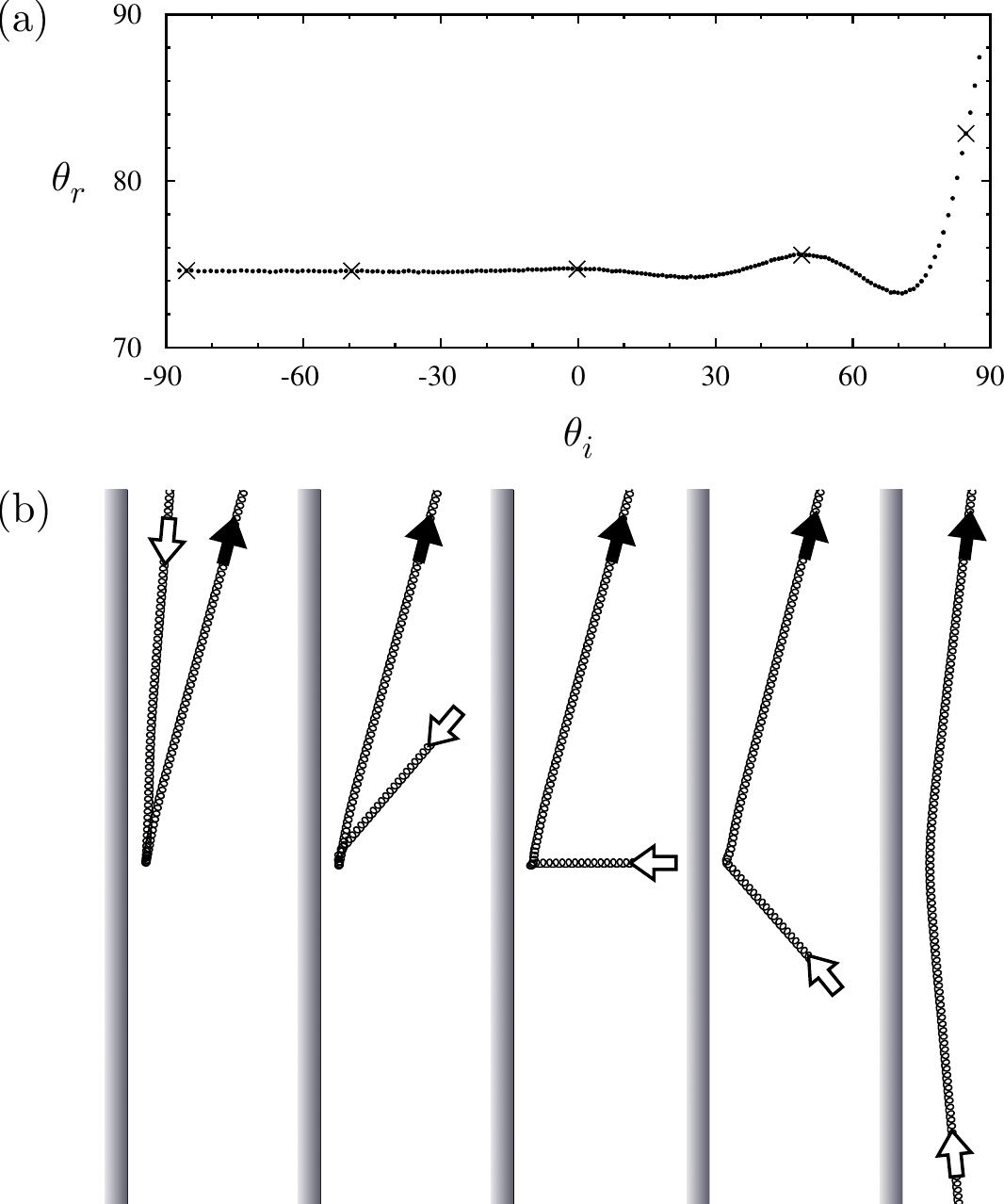}
\caption{
Illustrative results for reflection from a no-flux boundary, i.e.\ Neumann
boundary condition. (a) Reflected angle \thr\ versus incident angle \thi. (b)
Representative tip trajectories showing reflections at the incident angles
marked with crosses in (a). The reflected angle is nearly constant for the
full range of incident angles. The forcing amplitude is $A=0.072$. }
\label{fig:neumannBC} 
\end{figure}

\begin{figure}
\centering
\includegraphics[width=3.00in]{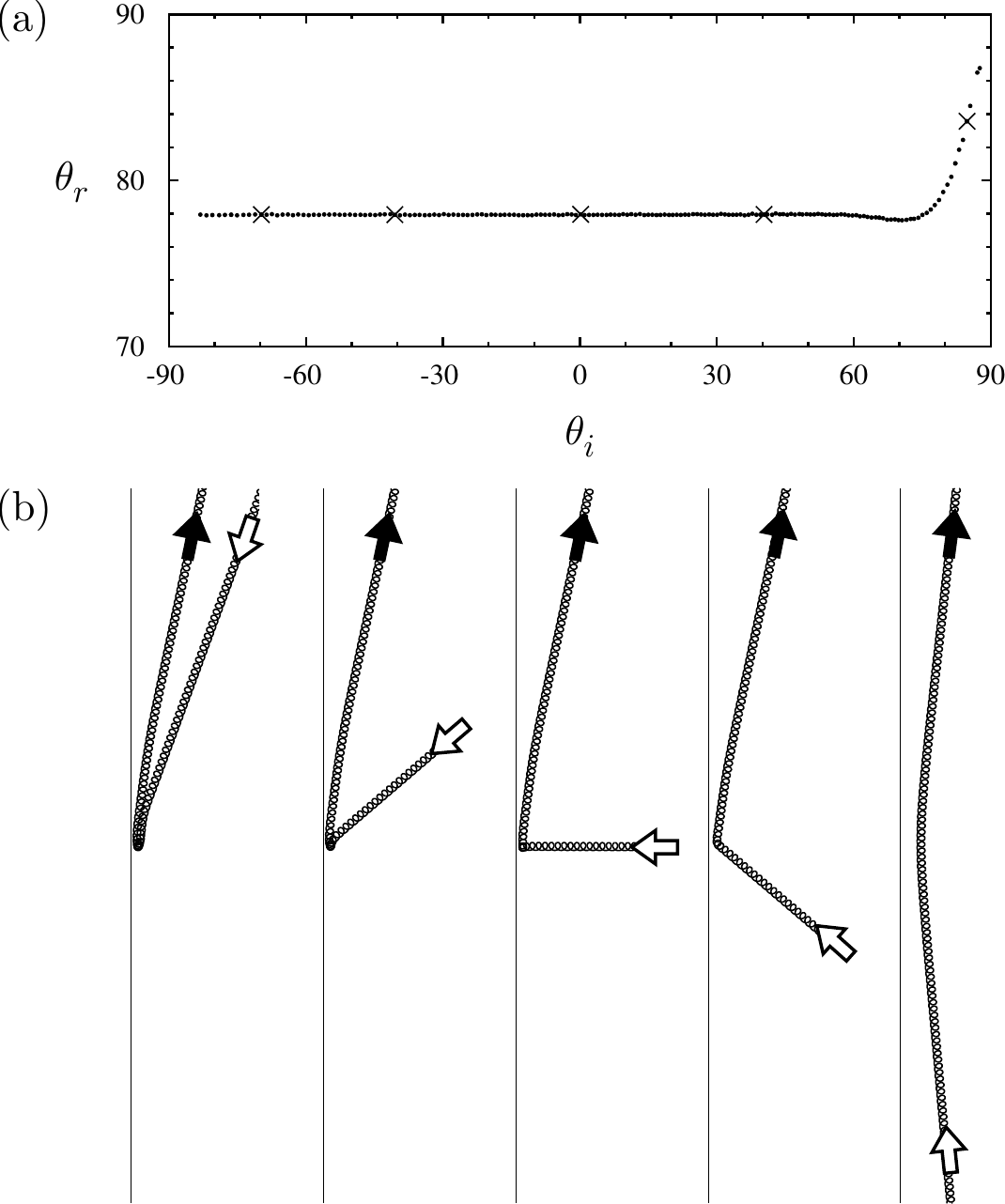}
\caption{
Illustrative results for reflection from a step boundary, i.e. a step
change in the excitability of the medium. (a) Reflected angle \thr\ versus
incident angle \thi. (b) Representative tip trajectories showing reflections
at the incident angles marked with crosses in (a). The reflected angle is nearly
constant for the full range of incident angles. The forcing amplitude is 
$A=0.072$. 
}
\label{fig:threshBC} 
\end{figure}

We begin with the small-core case already shown in Fig.~\ref{fig:intro}.
Figures~\ref{fig:neumannBC} and \ref{fig:threshBC}
illustrate the typical behavior we find in reflections from both types of
boundaries.  In both figures the upper plot shows measured reflected
angle \thr\ as a function of incident angle \thi\ over the full range of
incident angles. The lower plots show representative trajectories for specific
incident angles indicated.  
Here and throughout, the no-flux nature of the Neumann boundary is indicated
with shading ($x=0$ is the at the rightmost edge of the shading), while 
the step in excitability at a step boundary is indicated with sharp lines.
All parameters are the same for the two cases; they differ only in the type of
boundary that trajectories reflect from.

The reflections are 
far from specular. This is particularly striking
for $\thi < 0$ where the incoming and outgoing trajectories lie on the same
side of the normal. The reflected angle is nearly constant, independent of the
incident angle, except for incident angles close to $\thi=90^\circ$. There is
a slight variation in the reflected angle, seen as undulation in the upper
plots, but the amplitude of the variation is small. 

One can also observe in the lower plots that the point of closest approach is
also essentially independent of incident angle, except close to $\thi =
90^\circ$ where the distance grows. 
Spiral trajectories come much
closer to the step boundary than to the Neumann boundary. 

It is worth emphasizing that there is no effect of forcing phase in the results
presented in Figs.~\ref{fig:neumannBC} and \ref{fig:threshBC}. As the incident
angle is scanned, the phase of the spiral as it comes into interaction with
the boundaries will be different for different incident angles. While this
could have an effect on the reflected angle, 
we have verified that there is no such effect for the small-core cases 
we have studied, 
except at large forcing amplitudes near where spirals annihilate at the
boundary (discussed later).

\begin{figure}
\centering
\includegraphics[width=3.25in]{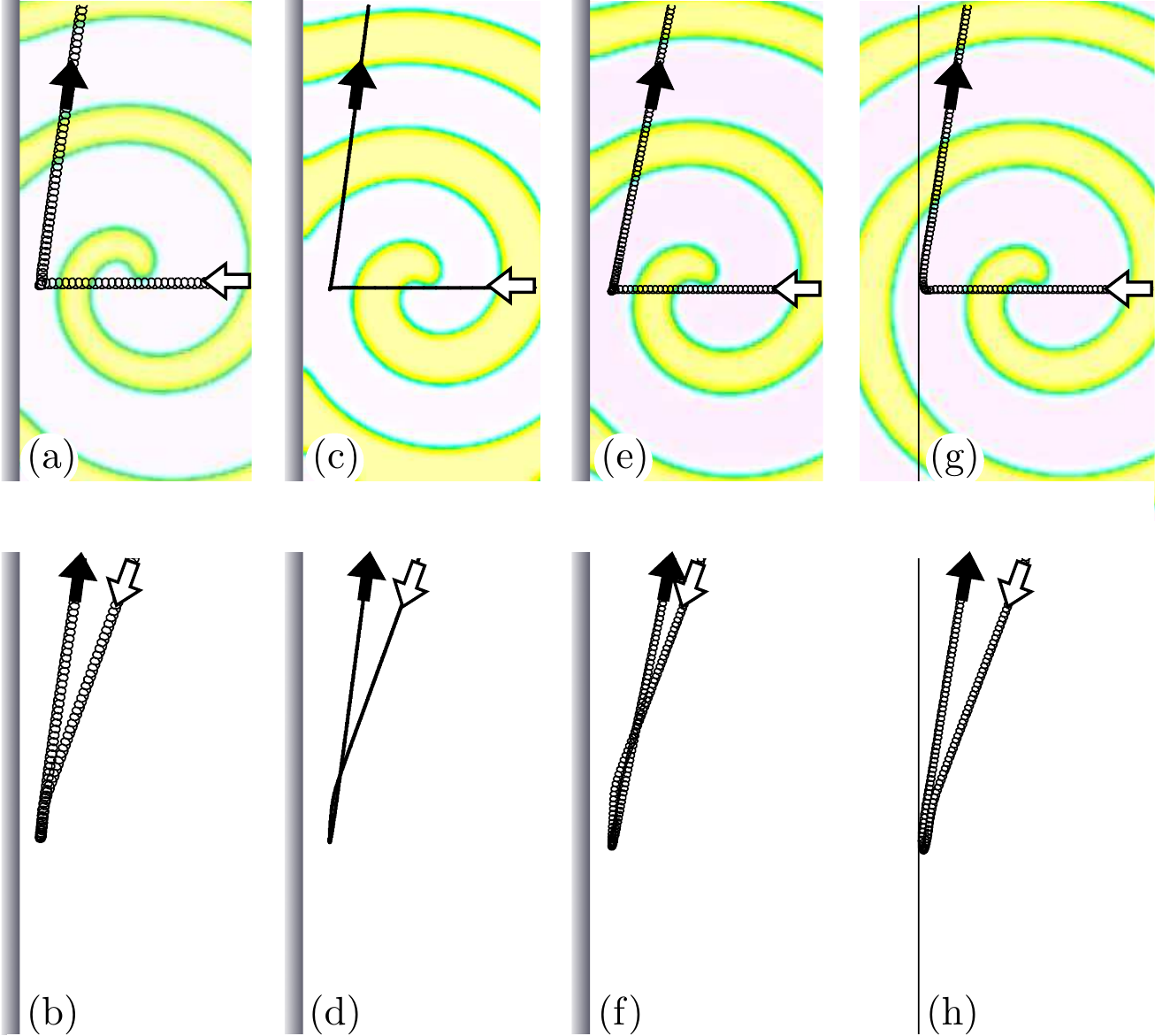}
\caption{
Illustration of the insensitivity of reflections throughout the small-core
region of parameter space. Upper plots show $\thi \approx 0^\circ$, including
faint visualization of the $u$-field at a particular time instance, while
lower plots show $\thi \approx -70^\circ$. The reflected angle is nearly
constant independently of incident angle, parameter values, and boundary type.
Model parameters span a substantial range of the non-meandering small-core
region: in (a) and (b) $a=0.7, b=0.01$; in (c) and (d) $a=0.95, b=0.01$; in
(e)-(h) $a=0.95, b=0.08$. Cases (g) and (h) are step boundary, the
others are all Neumann boundaries. $A=0.072$ throughout. }
\label{fig:params}
\end{figure}

While we have not conducted detailed studies at other parameter values, we
have explored the small-core region of parameter
space. Figure~\ref{fig:params} shows representative results at distant points
within the small-core region. The figure indicates not only a qualitative
robustness, but also a quantitative insensitivity to model parameter values
throughout the small-core region.
In each case the upper plot shows $\thi\approx 0^\circ$ while the lower plot
shows $\thi \approx -70^\circ$. The reflected angle varies by only a few
degrees throughout all cases shown in the figure. Case (a)-(b) is close to the
meander boundary while (c)-(d) is far from the meander boundary and
corresponds to a very small core. Cases (e)-(h) are relatively large
values of parameters $a$ and $b$, both with Neumann and step boundary
conditions.

In the step boundary case, there is also the effect of $\triangle b$ to
consider. Across a number of representative incident angles, we observed that
as $\triangle b$ is incremented from $0.025$ up to $0.05$, the closest
approaches of the spiral tips occur further from the boundary. We also find a
slight reduction in the angle of reflection. Decrementing $\triangle b$ has
the opposite effect. However, if $\triangle b$ is too small then the repulsive
effect at the boundary will be too small and spiral cores will cross the
boundary.

\begin{figure}
\centering
\includegraphics[width=3.25in]{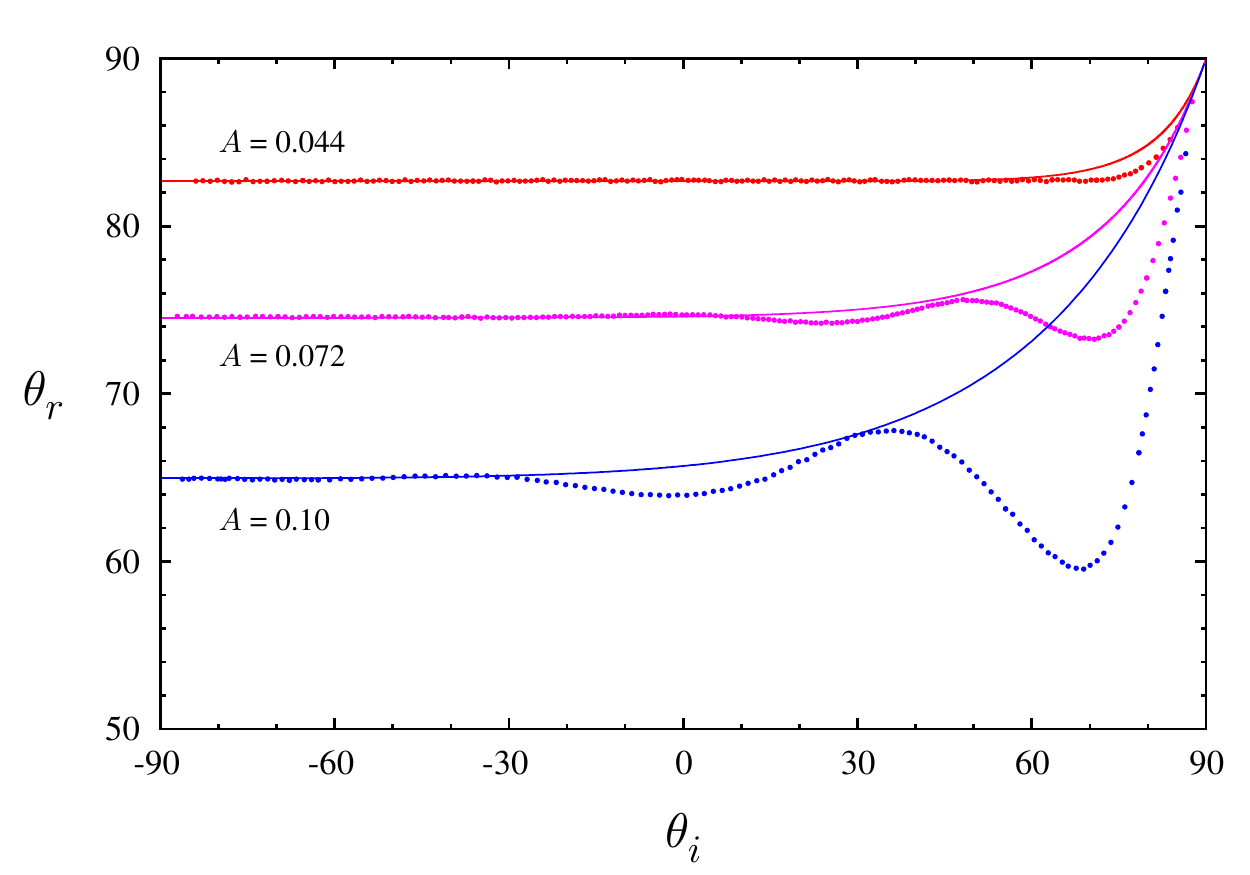}
\caption{
Effect of forcing amplitude on reflection of small-core spirals for the case
of a Neumann boundary.  Points are measured reflected angle as function of
incident angle at forcing amplitudes $A$ indicated. Solid curves are from 
Biktashev-Holden theory discussed in Sec.~\ref{sec:BH}. }
\label{fig:neumVarA} 
\end{figure}

\begin{figure}
\centering
\includegraphics[width=3.25in]{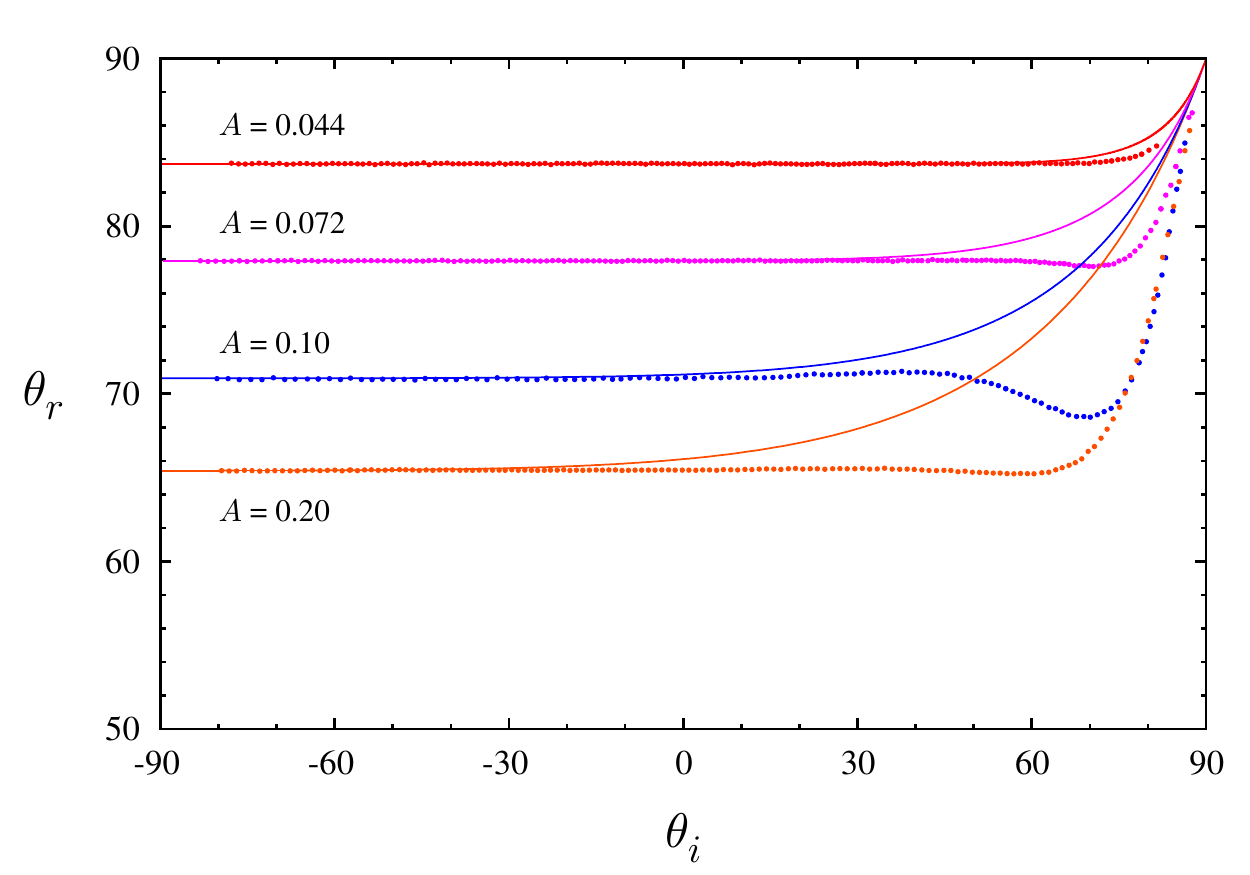}
\caption{
Effect of forcing amplitude on reflection of small-core spirals for the case
of a step boundary.  Points are measured reflected angle as function of
incident angle at forcing amplitudes $A$ indicated. Solid curves are from 
Biktashev-Holden theory discussed in Sec.~\ref{sec:BH}. } 
\label{fig:threshVarA}
\end{figure}

We have examined the effect of forcing amplitude
$A$. Figures \ref{fig:neumVarA} and \ref{fig:threshVarA} show reflected angle
as a function of incident angle for various values $A$ as indicated.  There is
a decrease in the reflected angle with increasing forcing amplitude, or
equivalently increasing drift speed. Generally there is also an increase in
the oscillations seen in the dependence of reflected angle on incident angle.
The solid curves are from the Biktashev-Holden theory discussed in
Sec.~\ref{sec:BH}.

\begin{figure}
\centering
\includegraphics[width=3.25in]{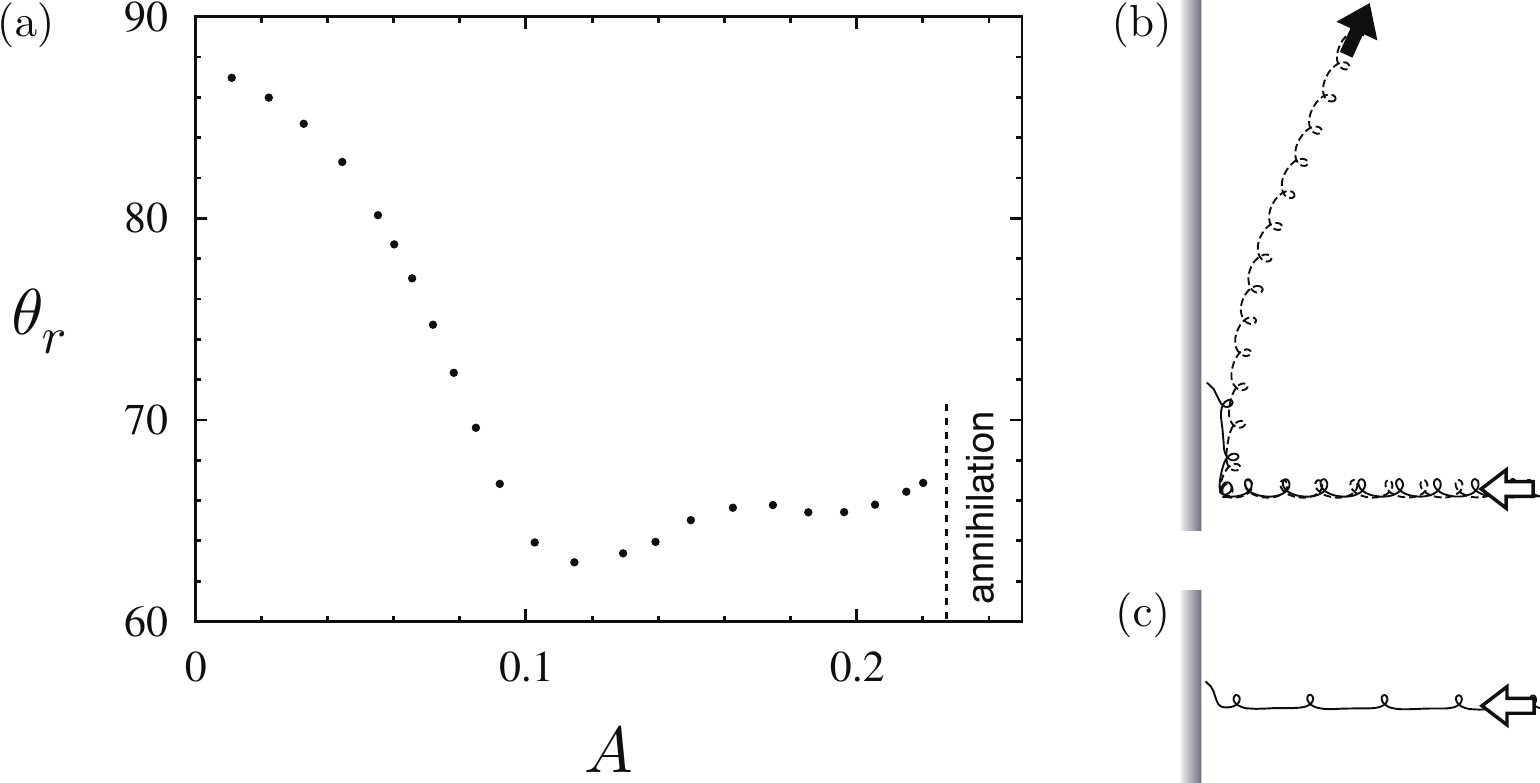}
\caption{
(a) Reflected angle as a function of forcing amplitude $A$ up to the point of
annihilation at a Neumann boundary for small-core spirals. The incident angle
is fixed at $\thi = 0^\circ$. 
(b) Tip trajectories a little below ($A = 0.215$) and a little above ($A =
0.235$) the forcing amplitude resulting in annihilation of the spiral at the
boundary.
(c) Tip trajectory at $A = 0.05$ showing annihilation at very large forcing
amplitude. 
}
\label{fig:annihil}
\end{figure}

For sufficiently large forcing amplitudes small-core spirals may be
annihilated as they drift into Neumann boundaries. In such cases no reflection
occurs.  We have not investigated this in detail as it is outside the main
focus of our study on reflections. Nevertheless, we have examined the effect of
increasing the forcing amplitude through the point of annihilation for the
case of a fixed incident angle $\thi = 0^\circ$.
The results are summarized in figure~\ref{fig:annihil}.
The reflected angle reaches a minimum for $A \simeq 0.11$, and thereafter
increases slightly, but does not vary by more than $4^\circ$ up to the
amplitude where annihilation occurs, $A \approx 0.225$, as indicated in  
figure~\ref{fig:annihil}(a).
The forcing amplitude at which annihilation first occurs is rather large in
that it corresponds to displacing the spiral considerably more than one
unforced core diameter per forcing period. Figure~\ref{fig:annihil}(b) shows
tip trajectories on either side of the amplitude where annihilation occurs,
while figure~\ref{fig:annihil}(c) shows annihilation at much larger forcing
amplitude.
We note that the exact amplitude at which annihilation first occurs depends
slightly on the rotational phase of the spiral as it approaches the
boundary. (Annihilation first occurs in the range $0.22 \lesssim A \lesssim
0.23$ depending on phase.)  Likewise, the spiral phase can affect the
reflected angle by nearly $1^\circ$ for $A \gtrsim 0.16$.
The influence of phase is nevertheless small for the small-core spirals. It
is, however, more pronounced in the large-core case which we shall now discuss.

\subsection{Large-core case}

\begin{figure}
\centering
\includegraphics[width=3.25in]{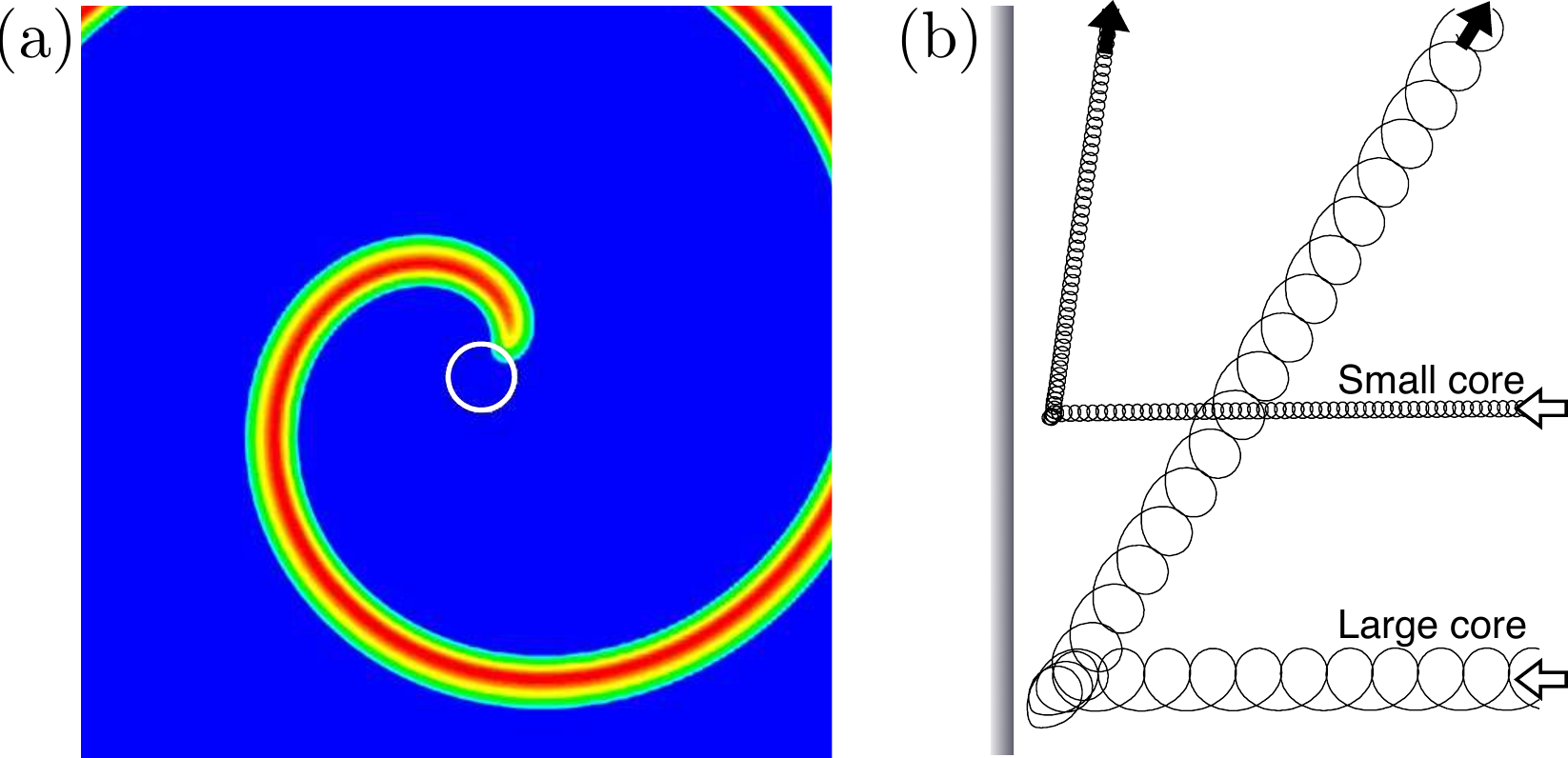}
\caption{Illustration of a large-core spiral wave. 
(a) A portion of a rotating spiral and corresponding tip trajectory in a
square region $40 \times 40$ space units [same size as
Fig.~\ref{fig:intro}(a)].
(b) Resonant forcing and reflection for a large-core spiral shown in
comparison to that of a small-core spiral. The forcing amplitude is $A=0.05$
in both cases.  }
\label{fig:large_core_intro}
\end{figure}

We now turn to the case where unforced spirals rotate around a relatively 
large core region
of unexcited medium. This case is illustrated in
Fig.~\ref{fig:large_core_intro}(a) where a rotating spiral wave and
corresponding tip trajectory are shown in a region of space the same size as
in Fig.~\ref{fig:intro}. The larger tip orbit and unexcited core, as well as
the longer spiral wavelength, in comparison with those of
Fig.~\ref{fig:intro}(a) are clearly evident. While such spirals occupy a
relatively narrow region of parameter space, they are nevertheless of some
interest because asymptotic treatments have some success in this
region~\cite{Mikhailov:1988,Hakim:1999} and because this is nearly the same
region of parameter space where wave-segments studies are
performed~\cite{Sakurai:2002,Zykov:2005,Steele:2008}.

Figure~\ref{fig:large_core_intro}(b) shows a typical case of non-specular
reflection for a large-core spiral compared with a small-core spiral forced at
the same amplitude. While many features are the same for the two cases,
large-core spirals are found often to reflect at smaller $\thr$ and moreover,
they can exhibit different qualitative phenomena.

\begin{figure}
\centering
\includegraphics[width=3.25in]{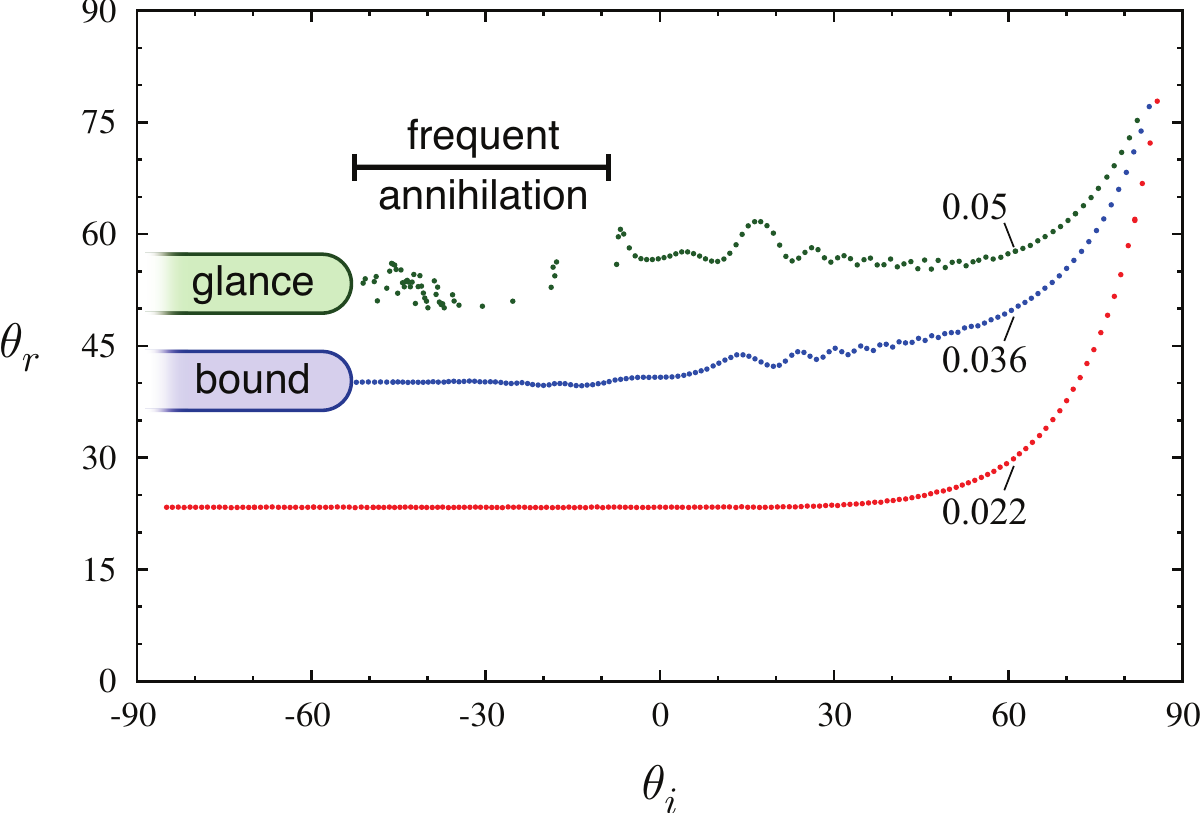}
\caption{Summary of results for large-core spirals. 
Reflected angle is plotted versus incident angle for three forcing amplitudes
as labeled. Neumann boundary conditions are used.  
For $A=0.05$ spirals are frequently annihilated at the boundary,
[Fig.~\ref{fig:large_core_paths}(d)], over the range of incident
angles indicated.
For $A=0.05$ and $\thi \lesssim -52^\circ$ trajectories glance from the
boundary [Fig.~\ref{fig:large_core_paths}(a)]. 
For $A=0.036$ and $\thi \lesssim -52^\circ$ trajectories become bound to the
boundary. [See text and Fig.~\ref{fig:large_core_paths}(b).]  
Wiggles are the effect of incident phase. 
}
\label{fig:large_core}
\end{figure}

\begin{figure}
\centering
\includegraphics[width=2.25in]{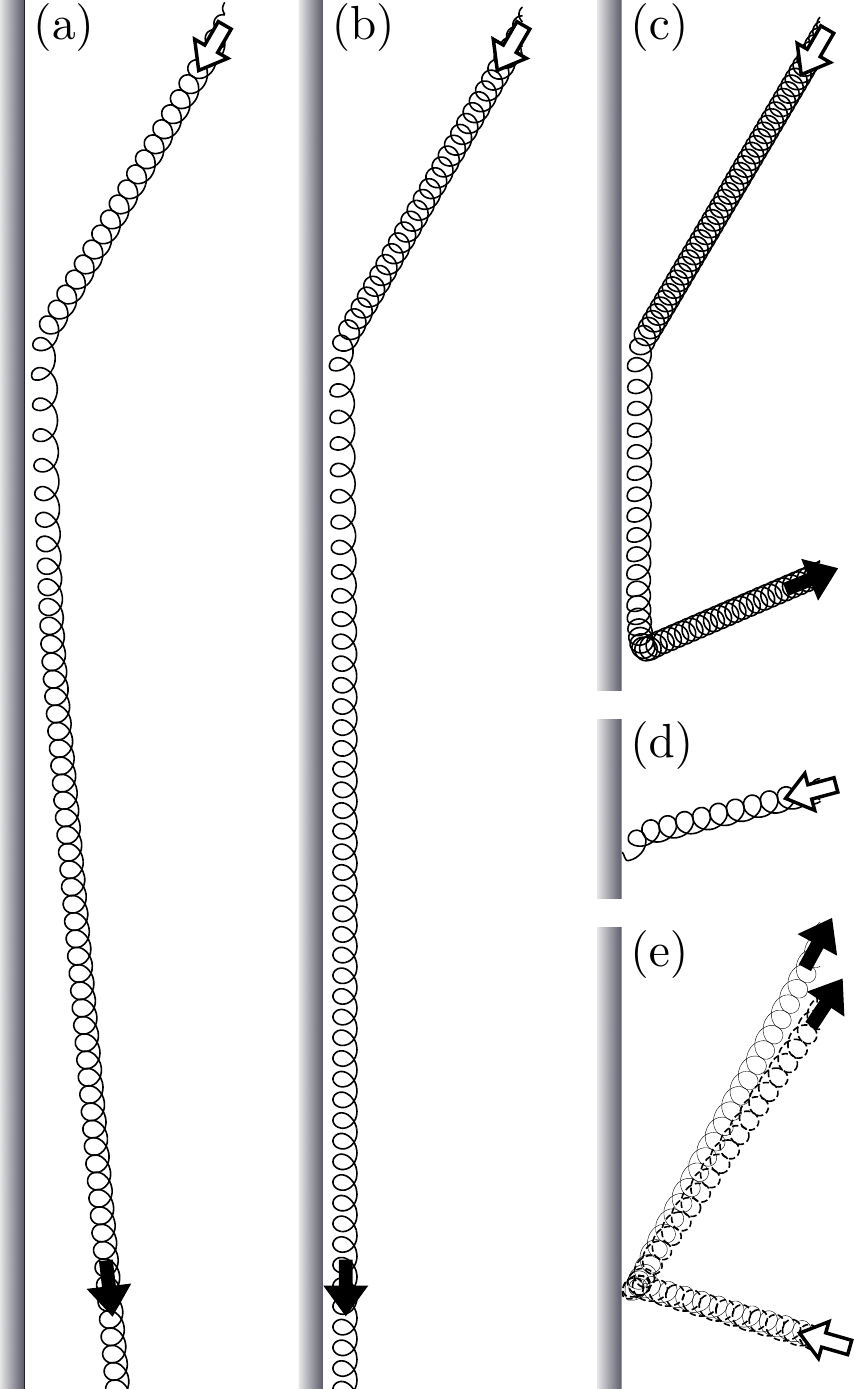}
\caption{Catalog of interesting trajectories for large-core spirals. 
(a)-(c) show impacts with $\thi \approx -60^\circ$ at different forcing
amplitudes. (a) With $A = 0.05$ the trajectory glances from the boundary and
moves off nearly parallel to it ($\thr \approx -85^\circ$). (b) With lower
amplitude $A = 0.036$, the trajectory becomes bound to the boundary. (c) With
yet lower amplitude $A = 0.022$, the trajectory hugs the boundary for a while
then leaves abruptly at an oblique angle ($\thr \approx 23^\circ$). In (d) the
incoming spiral with large forcing, $A = 0.05$, is annihilated at the
boundary. In (e) the effect of phase is seen with two approaching trajectories
shifted by half a core diameter. Otherwise the conditions are identical,
$A=0.05$.  The resulting reflected angles differ slightly. }
\label{fig:large_core_paths}
\end{figure}

Figures~\ref{fig:large_core} and \ref{fig:large_core_paths} summarize our
findings for large-core spirals. Reflected angle as a function of incident
angle for three forcing amplitudes is shown in Fig.~\ref{fig:large_core}. One
sees the overall feature, as with the small-core case, that reflected angle is
approximately constant over a large range of incident angles. This is
particularly true of low-amplitude forcing, $A=0.022$.  However, there are
also considerable differences with the small-core case. 

For large-core spirals the reflected angle increases with forcing
amplitude. This is opposite to what is found for small-core spirals in Figs.
\ref{fig:neumVarA} and \ref{fig:threshVarA}. 
Moreover, the reflected angles are noticeably smaller than for the small-core
case, as was already observed in Fig.~\ref{fig:large_core_intro}(b).

We now focus in more detail on what happens in various circumstances.
The left portion of Fig.~\ref{fig:large_core} indicates the different dynamics
we observe, depending on forcing amplitude, at large negative incident angles
($\thi \lesssim -52^\circ$), and
Figs.~\ref{fig:large_core_paths}(a)-\ref{fig:large_core_paths}(c) show
representative trajectories with $\thi \approx -60^\circ$. At $A=0.05$ trajectories
glance off the boundary. That is, they remain close for short while before
moving off with a well defined large negative reflected angle. The reflected
angle is nearly constant at $\thr \approx -85^\circ$ for incident angles
$\thi \lesssim -52^\circ$.
At $A=0.036$, $\thi \lesssim -52^\circ$, trajectories become bound to the
boundary and move parallel to it indefinitely.
In Fig.~\ref{fig:large_core_paths}(c), with $A=0.022$, one observes the
trajectory moving along the boundary for a distance before abruptly leaving the
boundary at a well-defined, relatively small positive reflected angle. 
This behavior is not restricted to $\thi \lesssim -52^\circ$ and is observed 
until $\thi \approx +20^\circ$. 
In fact, this type of reflection is also observed for the other two forcing
amplitudes studied for $\thi$ in a range above $-52^\circ$. For $A=0.036$ this
occurs until $\thi$ is approximately $-15^\circ$, while for $A=0.05$ this is
seen only until $\thi$ is about $-45^\circ$.

At the higher forcing amplitudes, as indicated for the case $A=0.05$ in
Fig.~\ref{fig:large_core}, large-core spirals are frequently annihilated when
they come into contact with the boundary. Figure~\ref{fig:large_core_paths}(d)
shows a typical example.  Whether or not a spiral is annihilated depends very
much on the spiral phase on close approach to the boundary.
The points shown in Fig.~\ref{fig:large_core} with $A=0.05$ are those where
the trajectory reflected; the absence of points indicates annihilation.
However, these results are for spirals all initiated a certain distance from
the boundary. Changing that distance would affect the spiral phase at close
approach and hence a different set of points would be obtained. Nevertheless,
the marked range of frequent annihilation is indicative of what occurs at this
forcing amplitude.

Finally, we address the wiggles in the reflected angle curves in
Fig.~\ref{fig:large_core}, most evident at large forcing amplitudes. These
wiggles are also due to the fact that the phase of spirals on close approach
varies with incident angle. 
Figure~\ref{fig:large_core_paths}(e) illustrates how the reflected angle
depends on phase by showing two trajectories shifted by half a core
diameter. This shifts the spiral phase upon approach to the boundary and
results in slightly different reflected angles. 
Rather than eliminating these wiggles by averaging over various initial
spiral distances, we leave them in as an indication of the variability due
to this effect. In general, 
reflections of large-core spirals are much more sensitive to phase 
than reflections of small-core spirals,
and one should understand that the data in Fig.~\ref{fig:large_core} will vary
slightly if similar cases are run with spirals initiated at different
distances from the boundary. 

While we have not studied the step boundary in detail for large-core spirals,
we have carried out a cursory investigation for such a boundary with
$\triangle b = 0.035$.  With the exception that there is no annihilation at
the step boundary, 
we observe qualitatively similar behavior to that just
presented for the Neumann case. Most notably we find both glancing and bound
trajectories.


\section{Discussion}

\subsection{Biktashev-Holden theory}
\label{sec:BH}

Many years ago Biktashev and Holden~\cite{bikt93,bikt95} carried out a study
very similar in spirit to that presented here.  Moreover, they understood that a
primary cause for the reflection from boundaries was the small changes in
spiral rotation frequency occurring as spiral cores came into interaction with
boundaries. Based on this they proposed an appealing simple model to describe
spiral reflections. The model is based on the assumption that both the
instantaneous drift speed normal to the boundary and spiral rotation frequency
are affected by interactions with a boundary, with the interactions decreasing
exponentially with distance from the boundary. While the actual interactions
between spiral and boundaries are now known to be more complex (see below), 
it is
worth investigating what these simple assumptions give. The beauty of the
simple model is that it can be solved to obtain a relationship between
reflected and incident angles, depending on only a single combination of
phenomenological parameters. (They called this combination $\theta$, but we
shall call it $p$. They also used different definitions for incident and
reflected angles.)

The model naturally predicts large ranges of approximately constant reflected
angle depending on the value of $p$. What is nice is that while fitting the
individual phenomenological parameters in their model would be difficult, it
is also unnecessary. The value of $p$ can be selected to match the plateau
value of $\thr$ observed in numerical simulations. Then the entire
relationship between $\thr$ and $\thi$ from the theory is uniquely determined.

Curves from the Biktashev-Holden theory are included in
Figs.~\ref{fig:neumVarA} and \ref{fig:threshVarA}. While there are obvious
limitations to the theory, it is nevertheless interesting to see that some of
the features are reproduced just from simple considerations. The theory would
be expected to work best where the drift speed is small: low amplitude
forcing. 
For the large-core spirals the theory does not apply and so the corresponding
curves are not shown in Fig.~\ref{fig:large_core}.

We do not intend here to propose a more accurate theory for the reflection of
drift trajectories. It is worth emphasizing, however, that in recent years
much has been understood about the interaction between spiral waves and
various symmetry breaking inhomogeneities
\cite{Biktasheva:2010,Biktashev:2010}. 
The key to this understanding has
been response functions, which are adjoint fields corresponding to symmetry
modes for spirals in homogeneous media
\cite{bikta03,Biktasheva:2009}. 
In principle, one could now obtain
rather accurate description of reflections using this approach, at least for
cases in which the boundary could be treated as a weak perturbation of a
homogeneous medium. 
This is beyond the scope of the present study and we leave this for future
research.

\subsection{Multiple reflections}

As noted in the introduction, Fig.~\ref{fig:intro}(d), when placed within a
square domain the trajectory of a drifting spiral will typically approach a
square, reflecting from each domain boundary such that $\thi + \thr =
90^\circ$. 
The reasons for this are simple (see for example Prati, {\em et
al.}\cite{Prati:2011}), but a brief analysis is useful, particularly for
understanding when square orbits become unstable.

\begin{figure}[h]
\centering
\includegraphics[width=2.5in]{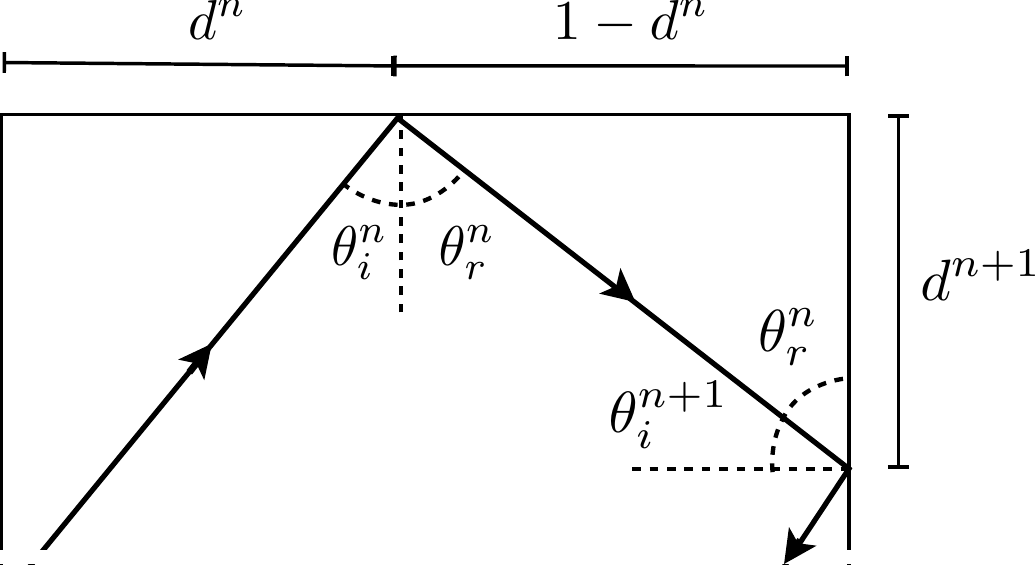}
\caption{
Sketch showing the geometry of multiple reflections in a portion of a square
box of normalised length. $d^n$ is the location, relative to the length of a side, of the $n^{th}$
reflection.
}
\label{fig:sketch}
\end{figure}

Figure~\ref{fig:sketch} shows the geometry of a consecutive pair of
reflections in the case where the reflected angle is larger than
$45^\circ$. In this case the path will necessarily strike consecutive sides of
the domain.
Consider first the path in terms of angles and let $\thi^n$ and $\thr^n$
denote, respectively, the $n^{th}$ incident and reflected angles, starting
from the initial reflection $\thi^0, \thr^0$. Trivially, the geometry of the
square domain dictates that $\thi^{n+1} + \thr^{n} = 90^\circ$. Then, if the
trajectory approaches an attracting path with constant angles,
$\lim_{n\to \infty} \thi^{n} = \thi^*$, $\lim_{n\to \infty} \thr^{n}
= \thr^*$, it must be that this path satisfies $\thi^* + \thr^* =
90^\circ$. That is, it must be a square or a rectangle.
Denoting the relationship between incident and reflected angle by $\thr
= \Theta(\thi)$, then a necessary condition for the square path to be
attracting is that $|\Theta^\prime(\thi^*)| < 1$. For the cases we
have studied this is true since $|\Theta^\prime(\thi^*)| \simeq 0$.

Turning now to the points at which the path strikes the edge of the domain, we
let $d^n$ denote the position of the $n^{th}$ reflection along a given side,
relative to the length of a side. One can easily see from the geometry that
$d^{n+1} = (1-d^n) \cot \thr^n$. Now, since $\thr \to \thr^*$, the fixed point
$d^*$ is given by $d^* = (1-d^*) \cot \thr^*$, or $d^* = 1/(1 + \tan \thr^*)$.
This corresponds to a square trajectory. 
For example, from Fig.~\ref{fig:neumVarA} with a forcing amplitude $A=0.072$
one can see that $\thr^*$ will necessarily be about $74^\circ$, giving
$d^* \approx 0.22$. These are the values seen in the simulation in
Fig.~\ref{fig:intro}(d).
A necessary condition for this fixed point to be stable is $|\cot \thr^*| <
1$. For small-core spirals $\thr^* > 45^\circ$, so $\cot \thr^*<1$,
and hence their square paths are stable. 

\begin{figure}[h]
\centering
\includegraphics[width=3.25in]{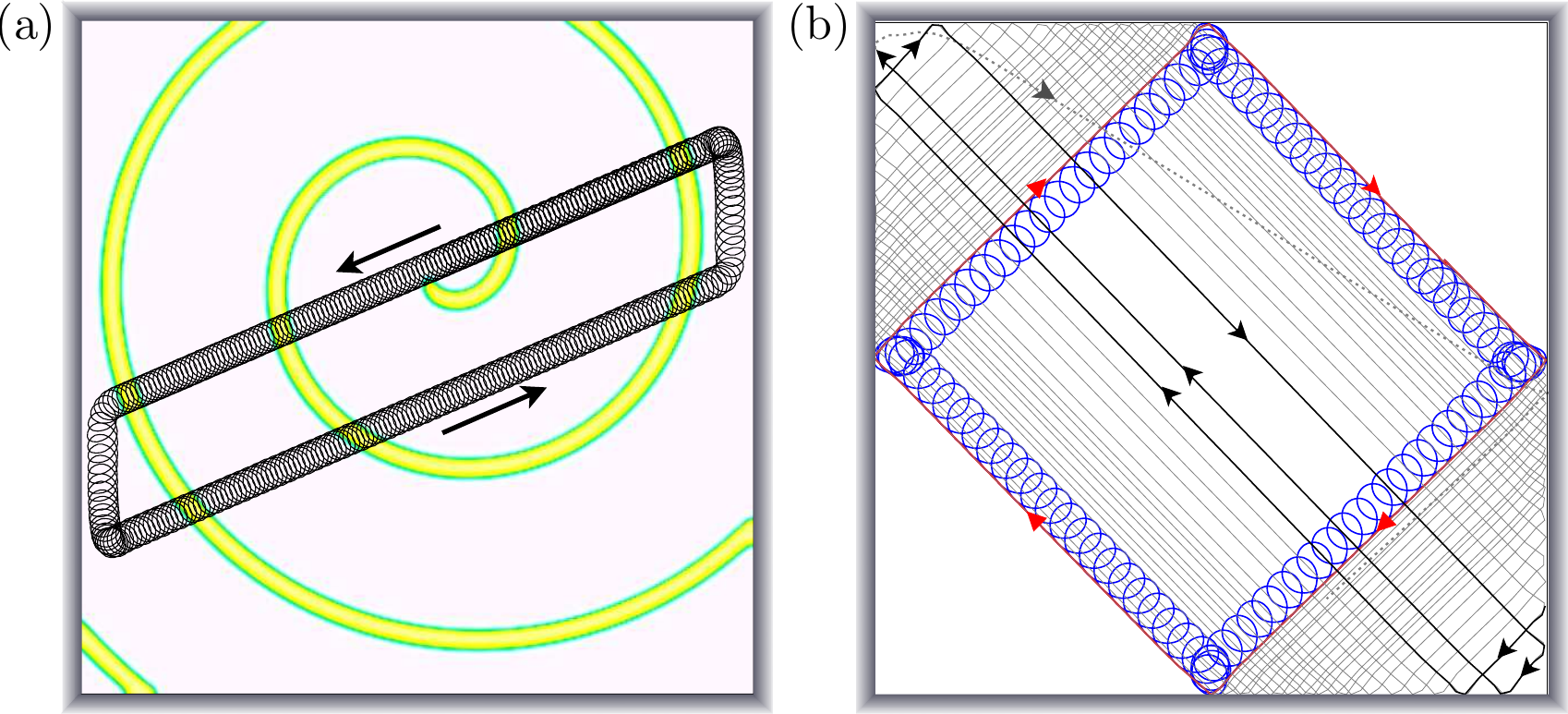}
\caption{Examples of non-square paths for large-core spirals. (a) 
$A=0.022$. The reflected angle is considerably smaller than $45^\circ$ and the
resulting trajectory bounces between opposite sides of the domain. The spiral
is shown faintly at one time instance.  (b) $A = 0.034$. The square trajectory
is unstable. For the first circuit around the nearly square path the full tip
trajectory is plotted. Subsequently, for clarity only, the tip path sampled
once per forcing period is shown. The trajectory collapses towards the
diagonal. The final portion of the trajectory before the spiral approaches the
corners is shown in bold. The spiral undergoes a complicated reflection from
the corner (gray, dotted).  
}
\label{fig:large_core_other}
\end{figure}

While square trajectories occur for small-core spirals, for large-core spirals
other trajectories are possible. Examples are shown in
Fig.~\ref{fig:large_core_other}. These occur when the reflected angle is
smaller than $45^\circ$.  (It is possible that $\thr < 45^\circ$ might occur for
small-core spirals in some regimes, although we have not observed them.) When
$\thr < 45^\circ$ it is not necessarily the case that trajectories will strike
consecutive sides of a square box. This is seen in
Fig.~\ref{fig:large_core_other}(a) where the spiral in reflects between
opposite sides of the domain. The reflections satisfy $\thr = -\thi$.

The more interesting case is when $\thr$ is only slightly less than $45^\circ$
as is seen in Fig.~\ref{fig:large_core_other}(b).  The square trajectory is
unstable.  While $|\Theta^\prime(\thi^*)| < 1$ and the angles converge quickly
to $\thi^n + \thr^{n} \simeq 90^\circ$, the equation $d^{n+1} =
(1-d^n) \cot \thr^n$ exhibits growing period-two oscillations for
$\cot \thr^n$ slightly larger than 1. Period-two oscillations in $d^n$ with
$\thi^n + \thr^{n} \simeq 90^\circ$ correspond to approximately rectangular
trajectories that approach a diagonal. This ultimately leads the spiral into a
corner of the domain where it may reflect in complicated manner. 

The analysis just presented should not be viewed as a model for spiral
trajectories.  Rather it just shows what global dynamics can be deduced simply
from a measured relationship between incident and reflected
angles. Essentially this same analysis appears as part of a study of cavity
solitions \cite{Prati:2011} which also undergo non-specular reflections from
walls and hence exhibit square orbits similar to figure~\ref{fig:intro}(d).
Our simple analysis should be contrasted with the situation for drops bouncing
on the surface of an oscillating liquid, so called {\em walkers}. Here physical
models of the liquid surface and drop bounces account for many varied features
of the system~\cite{prot06,coud06,Protiere:2008,fort10,Eddi:2011,Shirokoff:2012}. 
The corresponding theory for spiral waves would be that based on response
functions in which many details of drift trajectories can be
predicted \cite{Biktasheva:2010,Biktashev:2010}, although as of yet not
reflections from boundaries.
Memory effects are important for walkers because bouncing drops interact with
surface waves generated many oscillations in the past, and models necessarily
take this into account~\cite{coud06,fort10,Eddi:2011,Shirokoff:2012}. However, path
memory is absent from spiral waves in excitable media and this constitutes a
significant difference between the two systems. 

\subsection{Concluding remarks}

We have reported some quantitative and some qualitative features of
resonant-drift trajectories in excitable media. The main message is that
reflections are far from specular -- the reflected angle generally depends
only weakly on the incident angle and typically is nearly constant over a
substantial range of incident angles (particularly negative incident
angles). Biktashev-Holden theory~\cite{bikt93,bikt95} accounts for some of the
observed features, but a more detailed theory based on response
functions~\cite{bikta03,Biktasheva:2010,Biktashev:2010} is needed. 
We have seen that the behavior of large-core spirals is more varied than that
for small-core ones. 
Rather than simply reflecting from a boundary, large-core
spirals may sometimes become bound to, or glance from, or be annihilated at a
boundary, even at moderate forcing amplitudes.
Finally we have considered what can occur as spirals undergo multiple
reflections within a square domain, and in particular have shown that while
small-core spirals are observed to meet the conditions of stable square
trajectories, large-core spirals may fail to meet these conditions and
exhibit more interesting dynamics.

We motivated this study with a broader discussion of macroscopic systems with
wave-particle duality.  
A large number of analogues to quantum mechanical systems have been reported
for walkers on the surface of a vibrated liquid layer
\cite{coud05,coud052,prot06,coud06,eddi09,eddi092}. 
As far as we are aware, this is less the case for the propagating wave segments
studied by Showalter {\em et al.}\cite{Sakurai:2002,Zykov:2005,Steele:2008} or
the drifting spirals in excitable media considered here. (We examined briefly
small-core drift trajectories through a single slit, but did not observe
diffraction-like behavior.)
Nevertheless, for the reflection problem, spiral trajectories, propagating
wave segments, cavity solitons, and 
walkers all share the feature of non-specular
reflections~\cite{prot06,Steele:2008,Prati:2011,CouderFortPC} 
and as a result these
systems can show similar dynamics when undergoing multiple reflections within
a bounded region~\cite{prot06,eddi092,Steele:2008,Prati:2011,ShowalterPC}.
It will be of interest to make further quantitative comparisons between these
different systems in the future and to explore theoretical basis of this
behavior.

\section*{Appendix: Forcing parameter values}

In the appendix we report the exact values for the forcing parameters used in
the detailed quantitative incidence-reflection studies, since obtaining
high-precision values for resonant drift can be time consuming. The values
stated in the body of the paper are reported only to two significant figures.

\begin{table}[h]
\caption{\label{tab:resonant_params_small}Parameter values used to produce
  resonant (straight line) drift in the small-core case ($a=0.8$, $b =
  0.05$).}
\begin{ruledtabular}
\begin{tabular}{ ccc }
  $A$ & $\omega_f$ & $\triangle t$ \\
\hline
  0.044462 & 1.82 & 0.0187625 \\
  0.071868 & 1.792 & 0.0188508 \\
  0.102609 & 1.75 & 0.0188968 \\
  0.196132 & 1.63 & 0.0188957 \\
\end{tabular}
\end{ruledtabular}
\end{table}

\begin{table}[h]
\caption{\label{tab:resonant_params_large}Parameter values used to produce
  resonant (straight line) drift in the large-core case ($a=0.6$, $b =
  0.07$).}
\begin{ruledtabular}
\begin{tabular}{ ccc }
  $A$ & $\omega_f$ & $\triangle t$ \\
\hline
  0.022 & 1.025 & 0.0188035 \\
  0.035863 & 1.003 & 0.0187557 \\
  0.050144 & 0.989 & 0.0187961 \\
\end{tabular}
\end{ruledtabular}
\end{table}

\begin{acknowledgments}
We would like to thank Y.\ Couder, E. Fort, and K. Showalter for discussing
results prior to publication.
\end{acknowledgments}


%

\end{document}